# On-chip generation and guiding of quantum light from a site-controlled quantum dot.

Ayesha Jamil,[1] Joanna Skiba-Szymanska,[2] Sokratis Kalliakos,[2] Andre Schwagmann,[1,2] Martin B. Ward,[2] Yarden Brody,[1,2] David J. P. Ellis,[2] Ian Farrer,[1] Jonathan P. Griffiths,[1] Geb A. C. Jones,[1] David A. Ritchie,[1] and Andrew J. Shields[2*]

[1] *Cavendish Laboratory, University of Cambridge, J. J. Thomson Avenue, Cambridge CB3 0HE, United Kingdom.*

[2] *Cambridge Research Laboratory, Toshiba Research Europe Limited, 208 Science Park, Milton Road, Cambridge, CB4 0GZ, United Kingdom*

We demonstrate the emission and routing of single photons along a semiconductor chip originating from carrier recombination in an actively-positioned InAs quantum dot. Device–scale arrays of quantum dots are formed by a two–step regrowth process. We precisely locate the propagating region of a unidirectional photonic crystal waveguide with respect to the quantum dot nucleation site. Under pulsed optical excitation, the multiphoton emission probability from the waveguide's exit is $12 \pm 5$ % before any background correction. Our results are a major step towards the deterministic integration of a quantum emitter with the waveguiding components of photonic quantum circuits.

* email: andrew.shields@crl.toshiba.co.uk

The possibility of efficient linear optical quantum computing using single-photon (SP) sources, beamsplitters and SP detectors[1] has triggered a significant scientific effort in the field of quantum information science. While the use of photons as flying qubits has been implemented in a series of remarkable free-space or fiber-based proof-of-principle experiments,[2,3] the miniaturization of photonic quantum circuits down to the micro- or even nano-scale is essential for technological applications. A step towards this is the realization of quantum logic gates on silicon-based platforms,[4,5] where the logic operation is performed with single photons that are generated and detected externally. In most cases quantum light generation relies on the intense laser pumping of nonlinear crystals, which is inefficient and provides randomly-timed photons. For future scalable architectures, the integration of on-demand SP sources on a chip is highly desirable and marks a milestone towards the next generation of quantum photonic circuits for quantum information processing.

Semiconductor quantum dots (QDs) are considered as excellent candidates for integrated quantum light sources.[6] Efficient, on-demand SP emission resulting from spontaneous recombination of an exciton in a QD has been demonstrated for both optical and electrical carrier injection.[7-11] Coupling the QD with a microcavity can induce dramatic changes in the spontaneous emission rate as a result of the modified local density of optical states,[8,12-13] improving the efficiency of QD devices as on-demand SP sources. Prerequisites for efficient QD–cavity coupling are the spectral matching of the exciton emission with the cavity mode and the spatial overlap of the dipole (exciton) with the mode's electric field maximum. While the use of self-assembled QDs grown in the Stranski-Krastanov mode has been successful in demonstrating their potential as quantum light sources, their random position presents a major obstacle for their integration in future quantum photonic circuits and networks.

Control over the spatial position and arrangement of the quantum emitters on a chip can be achieved by the growth of site-controlled QDs (SCQDs). Several schemes to control the nucleation of QDs have been proposed and demonstrated, including the use of tetrahedral recesses[14,15] and nano-hole patterning.[16-19] QD emission from dry-etched pillars on quantum well substrates has been also reported.[20,21] SP emission from SCQDs has been shown by several groups using either optical excitation[22] or electrical injection.[23,24] Emission lines from SCQDs are typically broad, reflecting the formation of defects in the region surrounding the QD due to multi-step growth and ex-situ fabrication processes. However, linewidths as narrow as 7 µeV have been reported recently by using vertical stacking of QD layers, which has allowed the demonstration of indistinguishable single photons.[18] Up to now, active positioning of the quantum emitter has been highlighted in studies where the emission was directed *out of the chip plane*. The implementation of integrated SP emitters in quantum photonic networks requires the emission of quantum light *along the chip plane*. In this letter, we report the in-plane SP emission from a SCQD. For the transmission of quantum light along the chip, the QD is positioned at the centre of a photonic crystal waveguide (PCWG). The emission line from the SCQD was coupled to the slow-light mode of the PCWG, having a strong impact on the emission lifetime. Autocorrelation measurements of light collected from the exit of the waveguide show a very strong suppression of multi-photon events, both under pulsed and continuous-wave excitation.

The sample was grown by molecular beam epitaxy (MBE) in a two-stage process. First, a 900 nm thick sacrificial $Al_{0.7}Ga_{0.3}As$ layer was grown followed by the bottom half of the photonic crystal slab (105 nm GaAs). At this stage the wafer was removed from the growth chamber and the pits that serve as the nucleation sites for positioned dots were defined. Details on the wafer patterning and wafer preparation can be found elsewhere.[19] The wafer preparation process involves four de-oxidation steps. In the second growth stage 15 nm of GaAs re-growth buffer was grown followed by the InAs QDs. The amount of indium is determined by in-situ RHEED analysis during QD growth on an un-patterned test wafer and kept 8% below the dot formation threshold. This method assures that QDs are only formed in the predefined sites with single dot per site occupancy reaching 75%.[19] An atomic force microscope image of the SCQDs from an uncapped wafer is shown in Fig. 1 (a). The QDs were grown at 470° C and capped with 112 nm of GaAs after a 30 sec interrupt, resulting in a slab to support the photonic crystal structure of a 229 nm thickness with SCQDs right at the center of the slab for efficient light coupling. A schematic of the grown structure is shown in Fig. 2 (a).

PCWGs have been proven to be an efficient structure for the routing of quantum light along a semiconductor chip.[25,26] Placed in the waveguiding region, the emitter may take advantage of broadband coupling with the propagating mode[27] for efficient in-plane SP transfer. Moreover, slow-light effects[26] can alter the emission dynamics of the positioned dipole, improving the device performance considerably. To this end, we designed and fabricated photonic crystal W1 slab waveguides by omitting a line of air holes along the Γ–K direction in an equilateral triangular lattice geometry. The PCWGs were fabricated using standard electron beam lithography and dry etching processes.[26] The unidirectional devices are terminated at one end by a photonic crystal mirror and by a GaAs-air interface in the other. We align the center of the PCWG to the pre–defined position of the SCQD. Imaging analysis of several similar devices allowed us to determine the average offset of the center of the photonic crystal device with respect to the predefined etched pit to be 64 nm with a standard deviation of 21 nm.[28] According to our calculations for the fundamental mode in a photonic crystal L3 cavity with the same parameters as the one investigated here, the amplitude of the electric field drops by 50% at 70 nm from the center of the cavity where the maximum is. We may conclude that our control of position of the SCQD allows for efficient QD-cavity coupling in a relevant structure. We note that in experiments similar to the ones described below, we observed in-plane QD emission in more than half of the photonic crystal waveguide devices, which highlights the advantage of using SCQDs instead of self-assembled ones as integrated quantum emitters. In fact, the need of using wafers with ultra-low density of self-assembled QDs as quantum light sources limits the yield of similar operational devices on a chip to much less than 10%, according to our experience. A scanning electron microscope (SEM) image of several devices is shown in Fig. 1(b), with the center of the waveguiding region matching the SCQD position. In our experiment we optically excite the SCQD with a laser beam from the top of the device and probe the directed light emission along the PCWG by collecting from the PCWG exit, normal to the excitation laser.

Micro-photoluminescence experiments were carried out at cryogenic temperatures with the sample placed in a continuous-flow liquid helium cryostat. The spot from a 400 ps pulsed diode laser ($\lambda$ = 780 nm) was focused by a 50X microscope objective (numerical aperture NA=0.4) onto the top of the PCWG at a distance of ~ 8 µm from the edge of the sample. The laser spot has a diameter of ~ 1.5 µm. Light emitted from the end of the waveguide was collected by an identical microscope objective. The collected emission was dispersed by a single-grating spectrometer and recorded by a charge-coupled device[28].

Spectra from a PCWG with a lattice constant of $a$ = 246 nm and hole radius to lattice constant ratio $r/a$ = 0.345 at different temperatures are shown in Fig. 3 (a). At $T$ = 49 K the spectrum is dominated by two intense peaks at 940.2 nm and 942.2 nm, which are attributed to emission from the SCQD, probably from different charge configurations (it is not clear if they are related to charged or neutral states). In the following, we focus our analysis on the emission line at 940.2 nm. The small spectral shift with temperature of the high-energy peak at 939.3 nm compared to the QD emission line indicates that it is related to the photonic crystal structure. We have performed finite-difference time-domain calculations of the modal spectrum of a unidirectional waveguide[26] with the same design parameters, which are presented in Fig. 2 (b). Since we are collecting the light that is propagating along the waveguide, only the $y$-polarized modes are relevant to our experiment. This allows us to assign the observed high-energy peak to the fundamental $y$–polarized slow-light mode of the PCWG (at $\lambda$ = 921 nm). This claim is further supported by a study of L3 cavity devices in the center of the sample (out-of-plane experiments, not shown). A systematic change of the lattice constant in these devices allowed us to identify the observed fundamental and higher-order cavity modes when compared to calculations. The discrepancy between the theoretical and experimental values of the modes in these structures is comparable to the reported one for the in-plane device and is attributed to deviations from the intended values of parameters during the fabrication process (especially the hole radius $r$). Following a fitting procedure, the $Q$-factor is found to be ~ 760.

Temperature tuning allows us to couple the SCQD emission with the slow-light mode, as shown in Fig. 3 (a). A first indication of the coupling is the dramatic increase of the emission intensity, which is apparent at $T$ = 37 K. The coupling also has an impact on the spontaneous emission rate through the Purcell effect. The emission dynamics of the SCQD peak at different temperatures are shown in Fig. 3 (b). We observe a large decrease in the emission lifetime from the uncoupled case at $T = 49$ K (emission lifetime $\tau = 6.0$ ns) to the case where the slow-light mode and the SCQD emission peak are spectrally matched at $T = 37$ K ($\tau = 2.0$ ns) as a result of the cavity mode-SCQD coupling. This coupling can be seen as a proof of the efficient design and fabrication process of the device based on the controllable integration of a quantum emitter with a building block of a photonic quantum circuit.

Photon autocorrelation measurements were performed to assess the performance of our device as an in-plane SP source. Again, light was collected from the exit of the waveguide and a Hanbury-Brown and Twiss set-up[26] was used for measurements of the second–order correlation function $g^{(2)}(\tau)$: $g^{(2)}(\tau) = \langle I(t)I(t+\tau)\rangle / \langle I(t)\rangle \langle I(t)\rangle$, where $I(t)$ is the expectation value of the photon intensity at time $t$. Figure 4 (a) shows the coincidences histogram of the SCQD peak in resonance with the slow-light mode ($T = 37$ K) under pulsed excitation. Strong suppression of multiphoton events is observed at $\tau = 0$. The broad peaks reflect the long emission lifetimes and delayed emission from adjacent pulses overlaps with each individual peak. This phenomenon partially contributes to the coincidences that we observe at $\tau = 0$. Another factor that contributes to detected coincidences at $\tau = 0$ is our detector dark counts that are ~ 4% of the total detected counts. We performed a fitting procedure using Lorentzian functions without any background correction for the second-order correlation function. The results for different excitation powers are shown in the inset of Fig. 4 (a) with $g^{(2)}(0)$ values as low as $0.12 \pm 0.05$. Note that the multi-photon probability does not change dramatically even when the device is operated with excitation powers one order of magnitude higher. Similar behavior is observed under continuous-wave excitation ($\lambda$ = 632.8 nm). Operated at low excitation power ($P_{exc}$ = 450 nW), emission from the SCQD on resonance with the slow-light mode has $g^{(2)}(0) = 0.10 \pm 0.05$ (Fig. 4 (b)).

In conclusion, we have demonstrated the on-chip generation and propagation of single photons from a quantum emitter actively-positioned in the waveguiding region of a unidirectional photonic crystal waveguide. Coupling of the SCQD with the slow-light mode results in a substantial modification of the spontaneous emission rate. Strong in-plane suppression of multiphoton events is observed, with $g^{(2)}(0) < 0.15$. Our results show the potential of this system as an on-demand quantum light source for on-chip quantum processing. Embedded in more advanced structures[29] and employing modified overgrowth techniques,[18] one can anticipate high emission rates and long coherence times.

**Acknowledgements**

This work was partly supported by the EU through the Integrated Project Q-ESSENSE (contract no. FP7/2007–2013). A. J. acknowledges support from COMSATS Institute of Information Technology.

**Figure Captions**

Figure 1. (a) Atomic force microscope image from an uncapped wafer. QD formation is shown at pre-defined nucleation sites. (b) Scanning electron microscope image of several PCWGs with their waveguiding region centered at the QDs nucleation sites.

Figure 2. (Color online) (a) Schematic illustration of the grown layer sequence. The layer thicknesses are not to scale. (b) Calculated mode structure of a unidirectional PCWG. Red and black lines correspond to *x*-polarized and *y*-polarized modes, respectively. The grey areas to each side are used to define the spectral region of the photonic bandgap.

Figure 3. (Color online) (a) In-plane µ-PL spectra from a SCQD coupled to a PCWG at different temperatures. (b) Emission dynamics of the SCQD peak at different temperatures. The lines are exponential fits.

Figure 4. (Color online) (a) Autocorrelation histogram recorded with the SCQD peak on resonance with the slow-light mode (at $T = 37$ K) under pulsed excitation at 80 MHz. The extracted value of $g^{(2)}(0)$ is shown at the inset at four different excitation powers. (b) Second-order correlation function under continuous wave excitation. The thick red line is the outcome of a least squares fitting procedure.

**Figure 1**

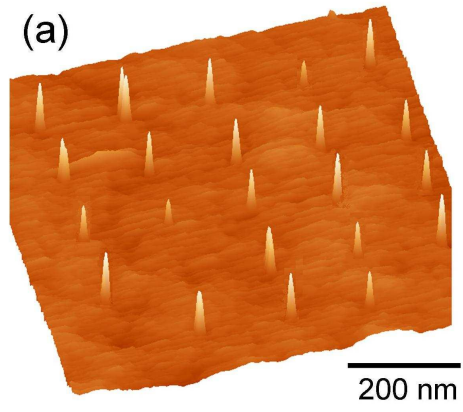 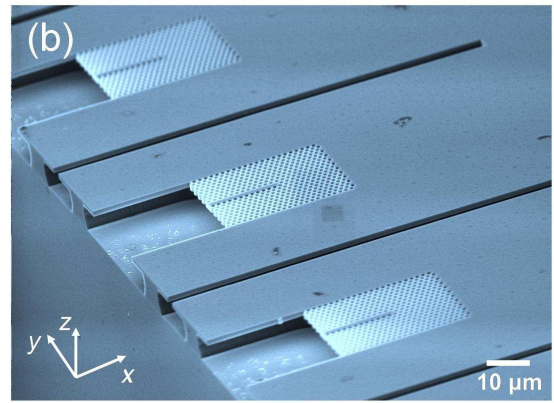

**Figure 2**

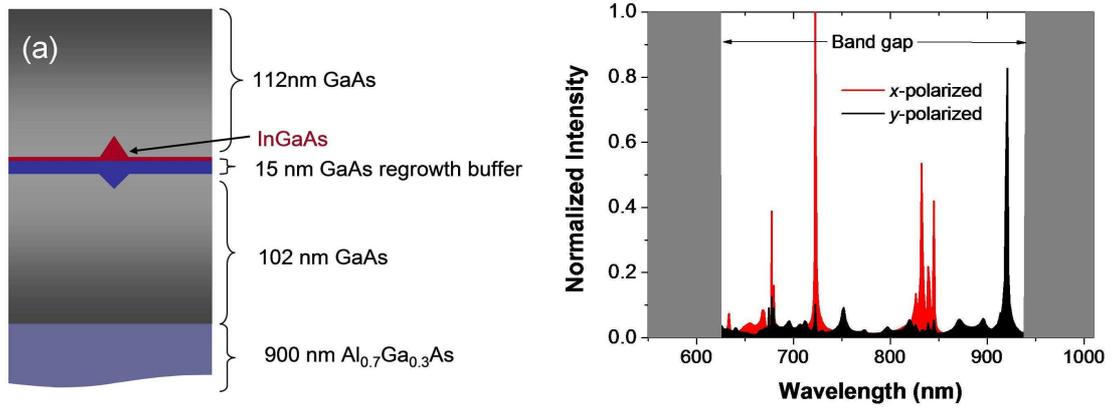

**Figure 3**

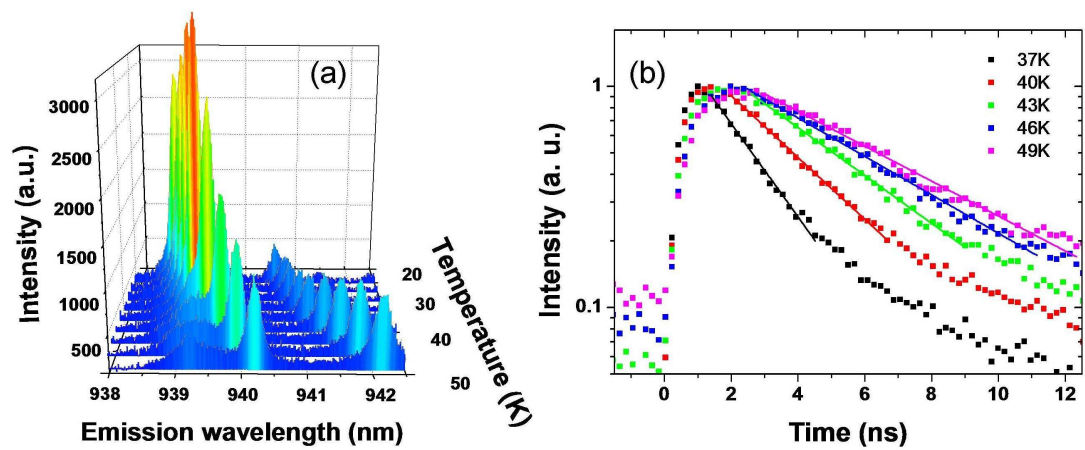

**Figure 4**

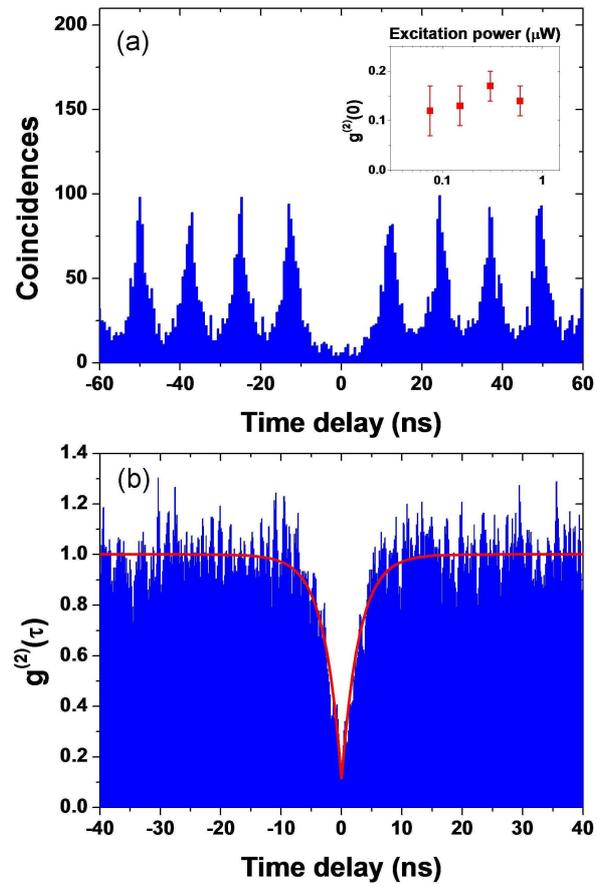